\documentclass[aps,prd,onecolumn,nofootinbib,superscriptaddress]{revtex4}
\pdfoutput=1
\usepackage{float}
\usepackage{graphicx}
\usepackage{amsmath}
\usepackage{amsfonts}
\usepackage{amssymb,ulem}
\usepackage{mathrsfs}
\usepackage{color}%
\usepackage{dcolumn}
\usepackage{subfigure}
\usepackage{pdfpages}
\usepackage{multirow}

\usepackage{MnSymbol,wasysym}
\usepackage{braket,diagbox}
\usepackage{eurosym}
\usepackage{euscript}
\usepackage[usenames,dvipsnames,svgnames]{xcolor}

\newcommand{\RNum}[1]{\uppercase\expandafter{\romannumeral #1\relax}}
\usepackage[colorlinks=true,linkcolor=blue,urlcolor=blue,filecolor=black,citecolor=red,
pdfstartview=FitV,pdftitle={},pdfsubject={},pdfkeywords={},pdfpagemode=None,bookmarksopen=true]{hyperref}

\begin{document}
\baselineskip=0.5 cm

\title{Detecting Lorentz-violation induced by a tensor field  with S-star's motion around Sgr~A*}

\author{Qi Qi}
\email{qiqiphy@163.com}
\affiliation{Center for Gravitation and Cosmology, College of Physical Science and Technology, Yangzhou University, Yangzhou, 225009, China}

\author{Yu Sang}
\email{sangyu@yzu.edu.cn}
\affiliation{Center for Gravitation and Cosmology, College of Physical Science and Technology, Yangzhou University, Yangzhou, 225009, China}

\author{Xiao-Mei Kuang}
\email{xmeikuang@yzu.edu.cn (corresponding author)}
\affiliation{Center for Gravitation and Cosmology, College of Physical Science and Technology, Yangzhou University, Yangzhou, 225009, China}

\date{\today}

\begin{abstract}
\baselineskip=0.5 cm
Testing Lorentz symmetry in strong gravitational fields provides a unique probe of extensions to standard model. The orbiting motions of the S-stars around the supermassive black hole Sgr~A* provide a natural laboratory for such tests. In this paper, we analyze the S2 orbital data focusing on a static and spherically symmetric black hole within Kalb--Ramond gravity, where the deviations from general relativity are encoded in a single Lorentz-violating parameter $\ell$ introduced by the Kalb--Ramond tensor field. Using a full 14-dimensional Markov Chain Monte Carlo analysis under uniform and Gaussian priors, we obtain $\ell = {1.60 \times 10^{-5}}^{+1.38 \times 10^{-4}}_{-1.76 \times 10^{-4}}  $ and $\ell =  {-1.02 \times 10^{-5}}^{+1.26 \times 10^{-4}}_{-1.19 \times 10^{-4}} $ at $1\sigma$ confidence level, respectively. These constraints are about three orders of magnitude tighter than those from Event Horizon Telescope imaging of Sgr~A*. We also perform MCMC simulation by fitting data of S38 and S55 stars, as well as their joint analysis. Our results show that the best fit values of $\ell$ in all simulations are always of $10^{-5}$ order, but S2 star provides the most stringent constraints on the parameters because  S2 star has higher precision observational data comparing to the fewer public data for other two stars.
\end{abstract}

\maketitle
\tableofcontents
\newpage

\section{Introduction}

While General Relativity (GR) has passed many verifications in the Solar System \cite{OLeary:2021twj,Pitjeva:2021hnc,GonzalezHernandez:2020evz,Carson:2020rea}, the environment around supermassive black holes (SMBHs) offers a much more extreme regime to search for potential deviations. For instance, the shadow images of M87* and Sagittarius~A* (Sgr~A*) provide direct ways to constrain the spacetime metric near the event horizon \cite{EventHorizonTelescope:2019dse, EventHorizonTelescope:2022wkp,EventHorizonTelescope:2021dqv, EventHorizonTelescope:2022xqj}. At the same time, future space-based gravitational wave detectors, such as the Laser Interferometer Space Antenna (LISA), the Taiji Program, and the Tianqin Program, are expected to provide more complementary precise measurement results about SMBHs \cite{LISA:2017pwj, Hu:2017mde, TianQin:2015yph}.
Besides the horizon scale, the Galactic Center provides another reliable method for testing GR through the motion of S-stars. These stars, which orbit around Sgr~A*, have been tracked with high precision for over three decades by the Keck Observatory and the Very Large Telescope (VLT) with the GRAVITY instrument \cite{Ghez:2003qj, Gillessen:2017jxc, GRAVITY:2021xju}. Among these, the S2 star is particularly useful because its close approach to the black hole makes it sensitive to relativistic corrections \cite{GRAVITY:2018ofz, GRAVITY:2020gka}. Current observations of S2 star have already verified GR predictions, including the gravitational redshift and the Schwarzschild precession \cite{Do:2019txf, GRAVITY:2023azi}. Although the current data remain consistent with GR, the uncertainties in the S2 star's orbit leave enough space for alternative models. Thus, using these orbital trajectories to place limits on additional degree of freedom in gravity theory were extensively studied, {such as analyzing the dark matter and dark energy distributions as well as the potential interactions \cite{Buckley:2017ijx,Nampalliwar:2021tyz,Tsai:2021irw,deLaurentis:2022oqa,Yuan:2022nmu,Bordoni:2025mli,Xamidov:2025prl,Tomaselli:2025zdo,Benisty:2021cmq,Benisty:2023clf}, and estimating parameters in modified gravities \cite{Shaymatov:2023jfa,DellaMonica:2021xcf,DellaMonica:2022eeg,Li:2022rjv,Yan:2022fkr,Zhang:2024fpm,deMartino:2021daj,Fernandez:2023kro,Li:2025qcv,GRAVITY:2023cjt,DellaMonica:2023dcw,DellaMonica:2023emp,Yan:2024bim,Li:2024tld,QiQi:2024dwc,Lafkih:2024qva,Yao:2025xeq,Xamidov:2025oqx,QiQi:2026pnb,Benisty:2023qcv}, and so on.}

One interesting of those extensions from GR is mainly driven by the possibility that Lorentz symmetry might be violated in a fundamental theory of gravity. Theories such as string theory and loop quantum gravity suggest that Lorentz symmetry breaking (LSB) could occur at the Planck scale and manifest as small corrections in the gravitational field \cite{Kostelecky:1988zi, Amelino-Camelia:2008aez, Bluhm:2004ep}. To effectively analyze these corrections, the Standard-Model Extension (SME) provides a comprehensive framework \cite{Colladay:1998fq, Kostelecky:2003fs}. Within this framework, the theory with Kalb--Ramond field, an antisymmetric tensor field, has gained significant attention because it arises naturally in the effective action of string theory \cite{Yang:2023wtu, Chakraborty:2014fva, Lessa:2019bgi}. Investigating the black hole solution in the gravity theory with the Kalb--Ramond field, enable one to translate these high-energy physics ideas into specific predictions that can be tested with astronomical data \cite{Deng:2025atg, Shi:2025rfq, Razzaq:2025rjh}.
In this scenario, the spontaneous breaking of Lorentz symmetry is triggered by a nonminimal coupling between the Kalb--Ramond field $B_{\mu\nu}$ and  gravitational field, which can develop a nonzero vacuum expectation value (VEV) $\langle B_{\mu\nu} \rangle = b_{\mu\nu}$, effectively setting a preferred direction in spacetime \cite{Altschul:2009ae}. The dynamics of this system in asymptotically flat spacetime are described by the action \cite{Kostelecky:1988zi, Kostelecky:1989jw, Bluhm:2004ep, Yang:2023wtu}
\begin{equation} \label{eq:Action}
S_{\text{KR}}=\int{d^4x} \mathscr{L}_{\text{KR}} =\frac{1}{2\kappa}\int{d^4x} \sqrt{-g}[ R+\xi B_{\mu\nu}B^{\mu\nu} R-\frac{1}{6} H_{\mu\nu\rho}H^{\mu\nu\rho}- V(B_{\mu\nu}B^{\mu\nu})],
\end{equation}
where $\xi$ represents the coupling strength between the Kalb--Ramond field and the spacetime curvature and $\kappa=8\pi G$ denotes the  gravitational constant. The field strength is defined as $H_{\mu\nu\rho} = \partial_{[\mu} B_{\nu\rho]}$, which provides the kinetic part of the Kalb--Ramond field. To ensure the theory remains invariant under local Lorentz transformations of observer, the self-interaction potential $V$ is designed to depend on the scalar $B_{\mu\nu} B^{\mu\nu}$. It is the potential, which leads to the spontaneous symmetry breaking happening, and forces the field to settle in a VEV given by $\langle B_{\mu\nu} \rangle = b_{\mu\nu}$. Here $b_{\mu\nu}$ is a tensor with a constant norm, i.e., $b_{\mu\nu} b^{\mu\nu} = \pm b^2$.

The derivation of exact black hole solutions under the Kalb--Ramond field, provides a foundation for investigating LSB in the strong field in this scenario \cite{Yang:2023wtu, Lessa:2019bgi,Ding:2019mal,Chakraborty:2014fva}. A static and spherically symmetric Schwarzschild-like metric  was obtained by assuming a specific radial configuration for the background Kalb--Ramond field, which is described by the metric \cite{Yang:2023wtu}
\begin{eqnarray}
ds^{2} &=&-g_{tt}dt^2+g_{rr}dr^2+r^2(d\theta^2+\sin^2\theta d\phi^2)\label{eq:general metric}\\
 &=&-\Big(\frac{1}{1-\ell}-\frac{2M}{r}\Big)dt^{2} + \Big(\frac{1}{1-\ell}-\frac{2M}{r}\Big)^{-1}dr^{2} + r^{2}(d\theta^{2}+\sin^{2}\theta\,d\phi^{2}).\label{eq:kr_metric_sec2}
\end{eqnarray}
The difference between the above metric and Schwarzschild black hole in GR are controlled by the dimensionless parameter $\ell \equiv \xi b^2$, which represents the combined effect of the nonminimal coupling and the VEV of the Kalb--Ramond field. In this geometry, the LSB parameter $\ell$ acts as a scaling factor that modifies the radial and temporal components of the metric. This kind of change leads to an observable difference from the Schwarzschild geometry and could be detected through astronomical data. In order to ensure the metric is physically reasonable and maintain the $(- , +, +, +)$ signature, we require that $1 - \ell > 0$, or $\ell < 1$. In most applications, $\ell$ is treated as a small correction, $|\ell| \ll 1$, which keeps this model consistent with GR and enable one to look for some signs of LSB. Moreover, a growing work has investigated the influence of the Kalb--Ramond field on various physical phenomena in the vicinity of black holes. In particular, extensions of this framework have explored both the dynamically perturbative properties \cite{Gu:2025lyz,Moreira:2025onu,Deng:2025atg,Baruah:2025ifh} and the thermodynamics \cite{Mangut:2025gie,Liu:2025fxj,Sucu:2025lqa,Jumaniyozov:2025dyy} of black holes in Kalb--Ramond gravity. Beyond the background spacetime itself, the effects of the Kalb--Ramond field on the motion and propagation of particles and test fields near black holes have also been extensively studied \cite{Shi:2025rfq,AraujoFilho:2024ctw,Razzaq:2025rjh,AraujoFilho:2025fwd,Cordeiro:2025cfo,Jumaniyozov:2025lox,Fayyaz:2025kqd}. In addition, its imprints on black hole shadows and images have also attracted increasing attention \cite{Yang:2025byw,AraujoFilho:2025huk,Cordeiro:2025eox,Xu:2025iwg,Zeng:2025kyv,Alrebdi:2025kqb,Sekhmani:2025zji}. Gravitational waves provide another powerful and complementary avenue for probing the Kalb--Ramond field. Recent studies have applied this framework to gravitational-wave observations \cite{Lobos:2026ysx}, with particular emphasis on waveform modifications in extreme mass-ratio inspirals (EMRIs) \cite{Xia:2025yzg}. Collectively, these investigations tended to study the observable and traceable signatures of LSB that can be tested against potential observational data.

Testing Lorentz symmetry induced from the Kalb--Ramond field in the strong field near a black hole offers an essential complement to constraints obtained in the existed literature. The SMBH at the Galactic Centre, Sgr~A*, offers a natural laboratory for such investigations. In this paper, we consider the orbital motion of the S2 star as a probe for LSB. By using the Schwarzschild-like metric in  \eqref{eq:kr_metric_sec2}, we can see how the parameter $\ell$ modifies the effective gravitational potential. This change directly affects the relativistic precession of the star's orbit, providing a way to constrain the influence of the Kalb--Ramond field from the star's trajectory.

To get a clear picture of these effects, we first derive the geodesic equations and obtain an analytical expression for the pericenter shift, keeping terms to leading order in $\ell$. Then we compare these theoretical predictions against the latest astrometric and spectroscopic data for S2 star. Finally, to explore the parameter space and its correlations between theoretical predictions and observable data, we run a Markov Chain Monte Carlo (MCMC) analysis covering a 14-dimensional parameter space. By testing the model with uniform and Gaussian priors, we aim to find the allowed range for $\ell$ and check its consistency with GR.


The paper is structured as follows. We start in Section \ref{sec:metrics_geodesics} by reviewing the geodesic motion of massive particles and calculating the pericentre precession for the Schwarzschild-like black hole. In Section \ref{sec: precession orbit}, we introduce the astrometric and spectroscopic data used in this work, and then explain the orbital model used in our MCMC simulations. The results of the MCMC analysis, including the constraints obtained under both uniform and Gaussian priors, are presented and discussed in Section \ref{sec:MCMC}. Finally, we summarize our main findings and offer some concluding remarks in Section \ref{sec:conclusion}.

\section{Geodesic motion and leading-order pericentre precession}
\label{sec:metrics_geodesics}

In this section, we examine the geodesic equations for timelike test particles to evaluate how the Lorentz-violating parameter $\ell$ affects the pericentre precession. The derivation follows the general method presented in \cite{Casana:2017jkc}, using geometric units $G=c=1$. Due to the spherical symmetry of the spacetime \eqref{eq:kr_metric_sec2}, we restrict our attention to the equatorial plane ($\theta = \pi/2$). The particle dynamics are described by the Lagrangian $\mathscr{L}=\tfrac12 g_{\mu\nu}\dot x^\mu\dot x^\nu$, where the overdot denotes differentiation with respect to the proper time $\tau$. Due to the symmetries of the metric, we can identify two conserved quantities, the energy $E$ and the angular momentum $L$ per unit rest mass,
\begin{align}
E &\equiv -g_{tt}\dot t, \label{eq:E_def_sec2}\\
L &\equiv g_{\phi\phi}\dot\phi = r^{2}\dot\phi. \label{eq:L_def_sec2}
\end{align}
Using the normalization $g_{\mu\nu}\dot x^\mu\dot x^\nu = -1$, the radial motion of a timelike particle is calculated by
\begin{equation}
\label{eq:radial_tau_sec2}
g_{rr}\dot r^{2} = E^{2} - V_{\rm eff}(r),~~~~\text{with}~~~~
V_{\rm eff}(r) \equiv -g_{tt}\Big(1+\frac{L^{2}}{r^{2}}\Big).
\end{equation}
By relating the \eqref{eq:E_def_sec2}, \eqref{eq:L_def_sec2} and \eqref{eq:radial_tau_sec2}, the trajectory equation can be expressed as
\begin{equation}
\label{eq:radial_phi_sec2}
\left(\frac{dr}{d\phi}\right)^{2} = \frac{r^{4}}{L^{2}}\frac{1}{g_{rr}}\Big[E^{2}-V_{\rm eff}(r)\Big].
\end{equation}
Introducing a new radial coordinate $u \equiv 1/r$, the second-order differential equation for the orbit is obtained by differentiating \eqref{eq:radial_phi_sec2} with respect to $\phi$
\begin{equation}
\label{eq:second_order}
\frac{d^2u}{d\phi^2}=\frac{1}{2L^2}\Big[-\frac{1}{g^2_{rr}} \frac{dg_{rr}}{du}(E^2 -V_{\rm eff})+\frac{1}{g_{rr}} (-\frac{dV_{\rm eff}}{du}) \Big].
\end{equation}

\begin{figure} [ht]
{\centering
\includegraphics[width=3 in]{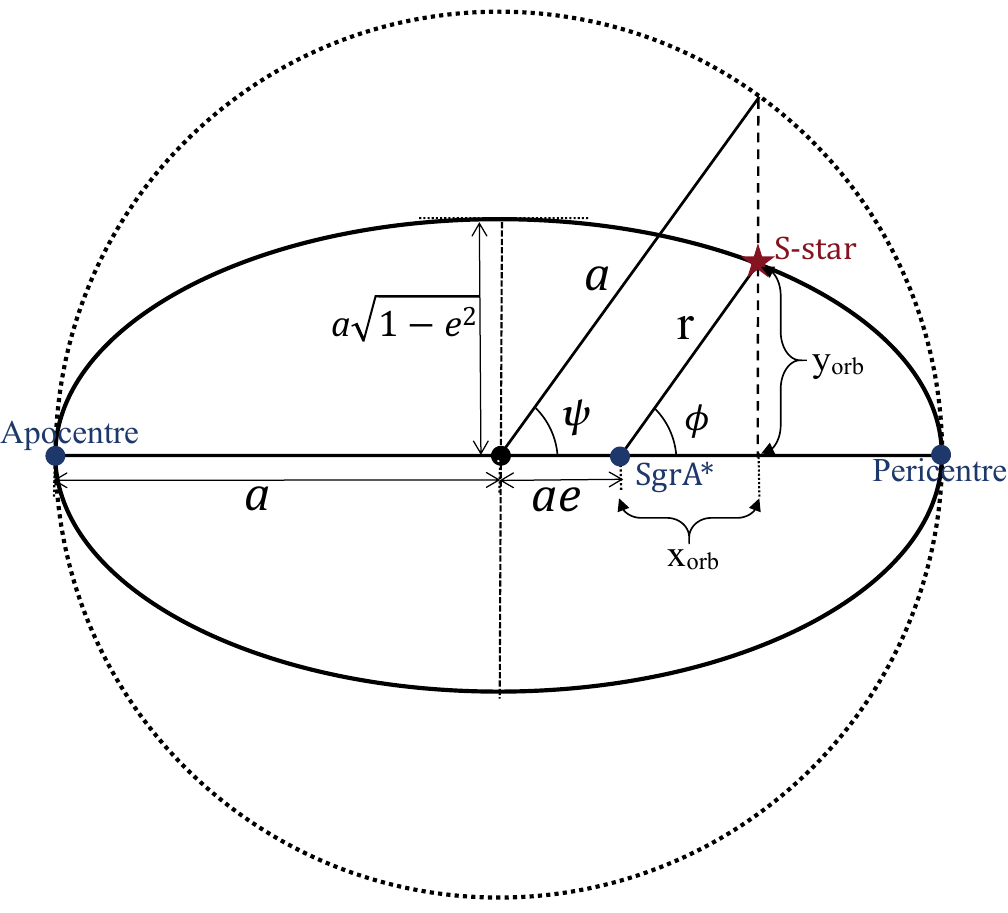} \hspace{0.5cm}
   \caption{ Orbit and orbital elements of the S2 star. The solid ellipse shows the trajectory, where $a$ and $e$ denote the semi-major axis and eccentricity. The angles $\psi$ and $\phi$ represent the eccentric and true anomalies, respectively, while $r$ is the distance from the S-star to the central black hole Sgr~A*. }   \label{fig:CCT}}
\end{figure}

Substituting the specific metric components from \eqref{eq:kr_metric_sec2} into the two conserved quantities \eqref{eq:E_def_sec2} and \eqref{eq:L_def_sec2} gives
\begin{equation}
E=\left(\frac{1}{1-\ell}-2Mu\right) \dot{t},~~L=\dot{\phi}/u^2 ,
\end{equation}
then the first order radial equation \eqref{eq:radial_phi_sec2} reduces to
\begin{equation}
\label{eq:kr_integ}
\Big(\frac{du}{d\phi}\Big)^2 = \frac{E^2}{L^2} - \frac{1}{L^2(1-\ell)} + \frac{2M}{L^2} u - \frac{u^2}{1-\ell} + 2M u^3.
\end{equation}
This expression is suitable for modeling the orbital motion of S-star around Sgr~A*, as illustrated in Fig.~\ref{fig:CCT}. The orbital trajectory is described by the semi-major axis $a$ and eccentricity $e$. Using the eccentric anomaly $\psi$, the motion of the S-star in the orbital plane can be parameterized as
\begin{equation}\label{eq:x(Psi)}
(\mathsf{x_{\mathrm{orb}}},\mathsf{y_{\mathrm{orb}}}) = (a (\cos \psi-e), a \sqrt{1-e^2} \sin \psi).
\end{equation}
The turning points of the motion are the pericentre and apocentre, which reach at $\psi=\phi=0$ and $\psi=\phi=\pi$, respectively. At these positions, the radial distance $r = \sqrt{\mathsf{x^2_{\mathrm{orb}}} + \mathsf{y^2_{\mathrm{orb}}}}$ simplifies to
\begin{align}\label{eq:r(Psi)}
r_p = a(1-e), ~~~~r_a = a(1+e),
\end{align}
where the subscripts $p$ and $a$ denote the values at the pericentre and apocentre, respectively.

At the orbital turning points, the radial velocity vanishes and $du/d\phi = 0$. This condition allows ~\eqref{eq:kr_integ} to be written as
\begin{equation}\label{eq:kr_integ-}
\frac{E^2}{L^2} - \frac{1}{L^2(1-\ell)} +\frac{2M}{L^2} u_{p,a} - \frac{u^2_{p,a}}{1-\ell} + 2Mu^3_{p,a} =0,
\end{equation}
where $u_a=1/r_a$ and $u_p=1/r_p$, respectively. By evaluating the difference between the two equations in~\eqref{eq:kr_integ-}, the energy $E$ can be eliminated, which facilitates the calculation of the angular momentum $L$. Considering that the orbital semi-major axis $a$ is much larger than the gravitational radius $M$, which means $M/a \ll 1$ \cite{Will:2014kxa, poisson2014gravity}, we focus on the leading-order Keplerian contribution to the angular momentum. Therefore, for $|\ell| \ll 1$, the expression of $L$ can be obtained as
\begin{equation} \label{eq:lequ}
L^2=Ma(1-e^2)(1-\ell)+ \mathcal{O}(\frac{M^2}{a^2}) \simeq Ma(1-e^2).
\end{equation}

The orbit of S-star can be further calculated by simplifying the second-order motion equation \eqref{eq:second_order} through a small-parameter approximation as
\begin{equation}
\label{eq:kr_integdouble}
\frac{d^2u}{d\phi^2} =\frac{1}{1-\ell} \Big[\frac{1}{a(1-e^2)}-u \Big] +3Mu^2 .
\end{equation}
To solve the analytical solution of equation \eqref{eq:kr_integdouble}, we employ a small post-Newtonian (PN) parameter defined as $\epsilon \equiv3M^2/L^2 \simeq3M/(a(1-e^2)) \ll1$. By expanding the radial coordinate in powers of the parameter $\epsilon$, $u \simeq u^{(0)} + \epsilon u^{(1)}$, and imposing the initial conditions $u(0)=u_p$ and $\dot{u}(0)=0$, the respective zeroth-order $u^{(0)}$ and first-order $u^{(1)}$ components are found to be
\begin{align}
u^{(0)} &=\Big(1-\ell \Big) \Big[1+e \cos(\frac{\phi}{\sqrt{1-\ell}}) \Big],\label{KR_integ}\\
u^{(1)} &\simeq(1-\ell)^3\Big[\Big(1+\frac{e^2}{2} \Big)-\frac{e^2}{6}\cos\Big(\frac{2 \phi}{\sqrt{1-\ell}}\Big)+\frac{e\phi}{\sqrt{1-\ell}}\sin\Big(\frac{\phi}{\sqrt{1-\ell}}\Big)\Big]. \label{eq:KR_integdouble}
\end{align}
Combining these terms and retaining only the leading-order term in $\epsilon$ allows the trajectory to be approximated as
\begin{equation}
\label{eq:KR_solution}
u \simeq u^{(0)}+\epsilon u^{(1)}\simeq (1-\ell)\Big[1+e\cos\Big[\Big(1-\epsilon (1-\ell)^2\Big)\frac{\phi}{\sqrt{1-\ell}}\Big]\Big] .
\end{equation}
The orbital period $\Phi$ is identified by the condition that the cosine value increases by $2\pi$ during one revolution. From \eqref{eq:KR_solution}, this requirement leads to the relation $(1-\epsilon(1-\ell)^2) \Phi / \sqrt{1-\ell} = 2 \pi n$, which yields
\begin{equation}
\label{KR_period}
\Phi=2\pi\sqrt{1-\ell}\Big(1+\epsilon (1-\ell)^2 \Big) =2\pi + \Delta\phi.
\end{equation}

By performing a Taylor expansion of $\Phi$ in terms of the small parameters $\ell$ and $\epsilon$, the anomalous precession per orbit, $\Delta\phi$, is
\begin{equation}
\label{eq:kr_orbit_period}
\Delta\phi \simeq2\pi \Big(1-\frac{\ell}{2}\Big) \Big(1+\epsilon \Big)-2\pi = \Delta\phi_{\rm GR}-\pi \ell .
\end{equation}
This result indicates that the Lorentz-violating parameter $\ell$ introduces a linear correction to the GR precession. In the absence of Lorentz violation, where $\ell=0$, the expression reduces to the Schwarzschild precession
\begin{equation} \label{eq:GRprece}
\Delta\phi_{\rm GR} = 2\pi\epsilon = \frac{6 \pi M}{a(1-e^2)}.
\end{equation}

%

\section{Data and data analysis of the S2 star}\label{sec: precession orbit}
The S2 star, a member of the nuclear stellar cluster, orbits around the supermassive black hole Sgr~A* at the Galactic Center. Its orbital properties are characterized by a period of approximately 16 years, a semi-major axis of about $970~\mathrm{AU}$, and a high eccentricity of $e \simeq 0.88$ \cite{Gillessen:2017jxc,Do:2019txf,GRAVITY:2018ofz}. Precise monitoring by the GRAVITY Collaboration over the past thirty years has yielded high-quality astrometric and spectroscopic records. These observations can be used to detect the relativistic phenomena, such as gravitational redshift and a pericentre precession of roughly $12'$ per period, both of them are consistent with GR.
The precision of the available S2 star's  data makes it possible to check for small potential deviations from GR near the supermassive black hole. In this work, we use the public observations of S2 to test the black hole \eqref{eq:kr_metric_sec2} that includes Lorentz-violating terms, which enables us to impose constraints on the parameters of Lorentz symmetry violation in SMBH scale within the proposed theoretical framework.

\begin{figure} [ht]
{\centering
\includegraphics[width=3.2in]{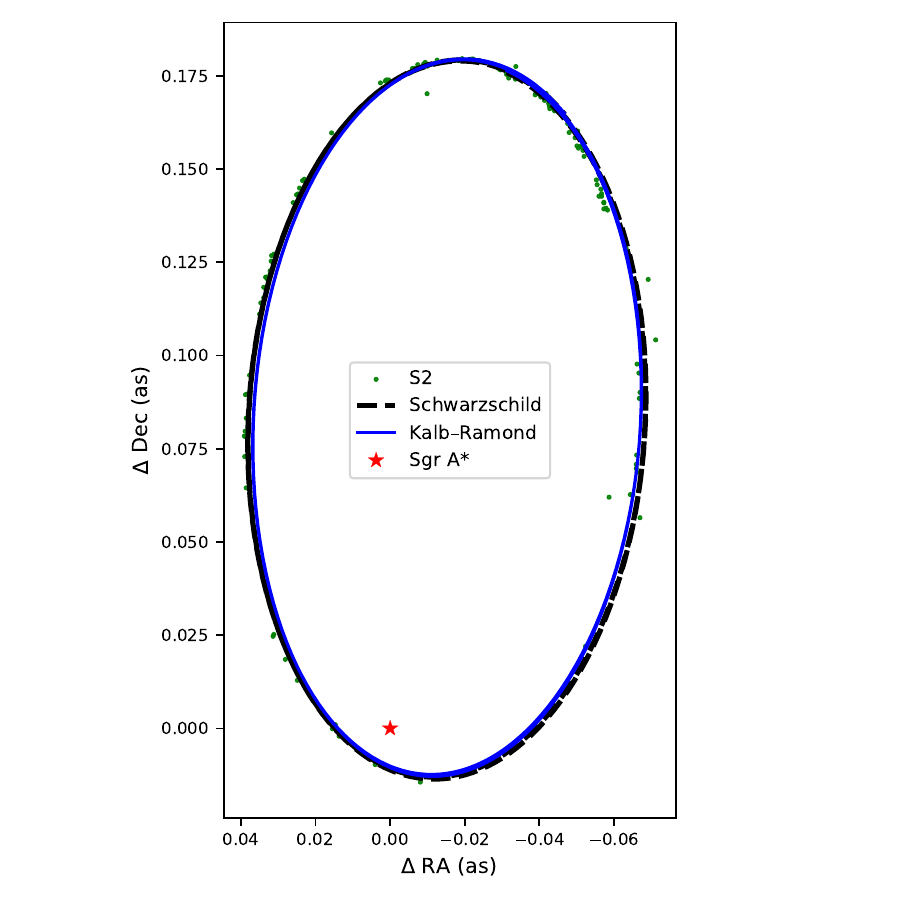}
\includegraphics[width=3.8in]{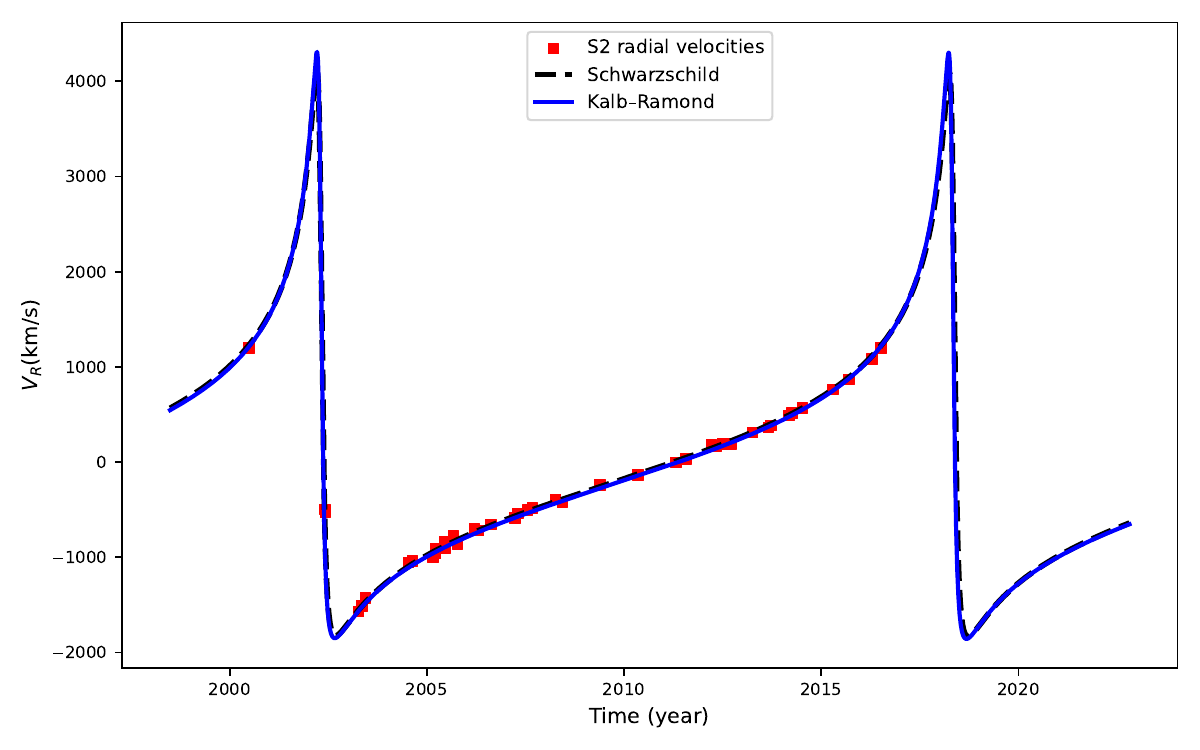}
   \caption{{\bf Left:} Comparison of the calculated orbit and the measured astrometric positions of the S2 star. The black dashed line represents the Schwarzschild black hole prediction, while the blue solid line shows the result for the Schwarzschild-like case. Green circles indicate the observed positions, and Sgr~A* is positioned at the origin, marked by a red pentagram. {\bf Right:} Radial velocity of the S2 star as a function of time. Similarly, The black dashed and blue solid curves correspond to the Schwarzschild and Schwarzschild-like predictions, respectively. Red squares show the observed radial velocity data. }   \label{fig:OBC}}
\end{figure}

\subsection{Dataset}\label{sec: observable dataset}
The analysis relies on public astrometric and spectroscopic data accumulated over several decades. The dataset is categorized into three primary types: astrometric positions, radial velocities, and the precession of the orbit. A brief summary of these data is provided below.

{\bf Astrometric positions ($\rm AP$):} We consider 145 measurements of the projected positions of the S2 star on the sky, covering the interval from 1992.224 to 2016.53 \cite{Gillessen:2017jxc}. These data were collected from different instruments: observations prior to 2002 were came from the SHARP speckle-imaging camera on the ESO New Technology Telescope (NTT), providing an accuracy of about $3.8~\mathrm{mas}$. Other measurements were taken with the NAOS+CONICA (NACO) adaptive-optics camera at the Very Large Telescope (VLT), which improved the precision to approximately $400~\mu\mathrm{as}$. These points are represented by the green dots in the left panel of Fig.~\ref{fig:OBC}.

{\bf Radial velocities ($V_R$):} This dataset includes 44 radial velocity points recorded between 2000.487 and 2016.519 \cite{Gillessen:2017jxc}. Similarly to the astrometric data, the Radial velocity measurements involve two main sources. Earlier data, before 2003,  were obtained via the NIRC2 spectrometer at the Keck Observatory, while more recent data were gathered by the Spectrograph for INtegral Field Observations in the Near Infrared (SINFONI), an adaptive-optics-assisted integral field spectrograph at the VLT. These measurement data are displayed as red squares in the right panel of Fig.~\ref{fig:OBC}.

{\bf Orbital precession ($\Delta \phi$):} We also consider the measurement of S2 star's pericentre advance as reported by the GRAVITY Collaboration \cite{GRAVITY:2020gka}. While the full astrometric dataset from GRAVITY is not yet public, the measured precession can be expressed as a ratio relative to the GR prediction
\begin{equation}\label{eq:deltaPHI}
f_{\rm sp}=\frac{\Delta\phi}{\Delta\phi_{\rm GR}} = 1.10 \pm 0.19 .
\end{equation}
This value is within $1\sigma$ of the GR prediction and deviates from a purely Newtonian model by more than $5\sigma$.

\subsection{Modeling the orbit with relativistic effects}
\begin{figure} [ht]
{\centering
\includegraphics[width=3 in]{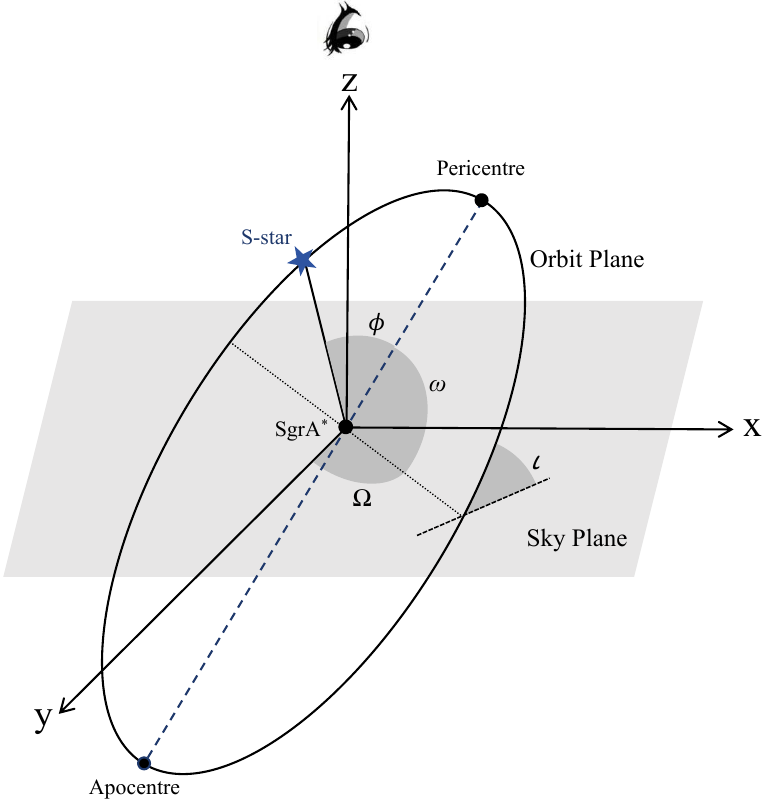} \hspace{0.5cm}
   \caption{ Geometry and Keplerian elements of the S2 star's orbit. The $Z$-axis is aligned with the observer's line of sight. The orbital orientation is defined by three angles: the longitude of the ascending node $\Omega$, the argument of pericentre $\omega$ (measured within the orbital plane from the ascending node), and the inclination $\iota$ between the orbital and sky planes. The star's position is indicated by the pentagram. }   \label{fig:CT}}
\end{figure}

To bridge the gap between theoretical predictions and astronomical observations, we numerically integrate the geodesic equations in the given black hole background \eqref{eq:kr_metric_sec2} to obtain the S2 star's trajectory in the orbital plane. However, astrometric data are recorded as projections onto the sky plane, the analysis requires both a geometric coordinate transformation and the consideration of observational corrections arising from the finite speed of light and reference frame instabilities. The geometric relationship between the orbital and sky planes is shown in Fig.~\ref{fig:CT}.

\textbf{(i) Coordinate transformation}
Since astrometric data are recorded as projections onto the sky plane (perpendicular to the observer's line of sight), we numerically integrate the geodesic equations in the orbital plane $(\mathsf{x_{\mathrm{orb}}}, \mathsf{y_{\mathrm{orb}}})$ and then transfer them to the observer's frame $(X, Y, Z)$. This transformation is governed by the inclination $\iota$, the longitude of the ascending node $\Omega$, and the argument of pericentre $\omega$. Using the Thiele-Innes constants, the sky-plane coordinates are expressed as
\begin{align}
X &= B \mathsf{x_{\mathrm{orb}}} + G \mathsf{y_{\mathrm{orb}}}, \label{eq:XX} \\
Y &= A \mathsf{x_{\mathrm{orb}}} + F \mathsf{y_{\mathrm{orb}}}, \label{eq:YY}\\
Z &= C \mathsf{x_{\mathrm{orb}}} + H \mathsf{y_{\mathrm{orb}}}. \label{eq:ZZ}
\end{align}
Here, the $Z$-axis points along the line of sight. The constants $A, B, C, F, G,$ and $H$ are defined following standard conventions \cite{Catanzarite:2010wa},
\begin{align}
A &= \cos\Omega\cos\omega - \sin\Omega\sin\omega\cos \iota, \
~~~B = \sin\Omega\cos\omega + \cos\Omega\sin\omega\cos \iota, \\
C &= -\sin\omega\sin \iota, \
~~~F = -\cos\Omega\sin\omega - \sin\Omega\cos\omega\cos \iota, \\
G &= -\sin\Omega\sin\omega + \cos\Omega\cos\omega\cos \iota, \
~~~H = -\cos\omega\sin \iota.
\end{align}

The velocities are transformed using the same logic:
\begin{align}
V_X &= B \mathsf{v_{x,\mathrm{orb}}} + G \mathsf{v_{y,\mathrm{orb}}}, \\
V_Y &= A \mathsf{v_{x,\mathrm{orb}}} + F \mathsf{v_{y,\mathrm{orb}}}, \\
V_Z &= -(C \mathsf{v_{x,\mathrm{orb}}} + H \mathsf{v_{y,\mathrm{orb}}}),
\end{align}
where the negative sign in $V_Z$ follows the convention that radial velocity is positive when the star moves away from the observer.

\textbf{(ii) Observational corrections}
Directly comparing the projected coordinates $(X, Y)$ with the data of S2 star from the GRAVITY or Keck groups requires addressing the R\o mer delay and the stability of the reference frame.

Firstly, we consider the reference frame offsets and drifts. The observational reference frame may have a offset or linear drift relative to the center black hole Sgr~A*. To account for these, we introduce four parameters, $\mathsf{x_0}, \mathsf{y_0}, \mathsf{v_{x0}}, \mathsf{v_{y0}}$ \cite{Do:2019txf}. Then, the sky-plane positions are corrected as
\begin{eqnarray}
X &=& X(t_{\rm em}) + \mathsf{x_0} + \mathsf{v_{x0}}(t_{\rm em}) (t_{\rm em}-t_{\rm refer}),\\
Y &=& Y(t_{\rm em}) + \mathsf{y_0} + \mathsf{v_{y0}}(t_{\rm em}) (t_{\rm em}-t_{\rm refer}).
\end{eqnarray}
In these relations, $t_{\text{em}}$ represents the time of photon emission, while $\mathsf{x_0}$ and $\mathsf{y_0}$ denote the constant coordinate offsets. The terms $\mathsf{v_{x0}}$ and $\mathsf{v_{y0}}$ describe the linear velocity drifts of the reference frame relative to the dynamical center \cite{Gillessen:2017jxc}. We adopt a reference epoch of $t_{\text{refer}} = 2009.2$, consistent with the value used in \cite{plewa2015pinpointing}.

Secondly, we consider the R\o mer time delay. The theoretical positions are calculated at the time of photon emission $t_{\text{em}}$, but astrometric measurements are recorded at the observation time $t_{\text{obs}}$. Due to the finite speed of light $c$ and the infinite distance between the star and the observer, these two time scales are related by \cite{Do:2019txf, GRAVITY:2018ofz},
\begin{equation}
t_{\rm em} = t_{\rm obs} - \frac{Z(t_{\rm em})}{c},
\end{equation}
where $Z(t_{\text{em}})$ is the line-of-sight coordinate calculated from Eq.~\eqref{eq:ZZ}. This correction ensures that each theoretical calculation point on the trajectory is compared to the correct observational point.

Finally, we account for the relativistic frequency shifts and radial velocity. The observed line-of-sight velocity $V_R$ is derived from the total frequency shift $\zeta$,
\begin{equation}
\zeta = \frac{\Delta\nu}{\nu} = \frac{\nu_{\rm em}-\nu_{\rm obs}}{\nu_{\rm obs}} = \frac{V_R}{c},
\end{equation}
here $\nu_{\rm em}$ and $\nu_{\rm obs}$ are the emission and observed photon frequencies, respectively. This shift consists of two main contributions: the special relativistic Doppler effect $\zeta_D$ due to the star's motion, and the gravitational redshift $\zeta_G$ caused by the spacetime curvature at the point of emission. These components are expressed as
\begin{align}
\zeta_D = \frac{\sqrt{1 - \mathsf{v_{\rm em}^2/c^2}}}{1 - \vec{n}\cdot \vec{\mathsf{v}}_{\rm em}/c}, ~~~
\zeta_G = \frac{1}{\sqrt{|g_{tt}|}},
\end{align}
where $\mathsf{v_{\text{em}}}$ is the velocity at the time of emission and $\vec{n}\cdot \vec{\mathsf{v}}_{\text{em}}$ is its component along the line of sight. Combining these gives the total shift $\zeta = \zeta_D \cdot \zeta_G - 1$. To match the actual spectroscopic measurements, we must also include a potential velocity offset $\mathsf{v_{z0}}$ of the central mass Sgr~A*. The final observed radial velocity is expressed as \cite{astro-ph/0612164}
\begin{equation}
\label{eq:total_radial_V}
V_R = c \cdot \zeta + \mathsf{v_{z0}}.
\end{equation}

Using the framework described above, we compare the orbital predictions of the Schwarzschild-like metric \eqref{eq:kr_metric_sec2} with those of the standard Schwarzschild black hole. The resulting trajectories and radial velocity curves are shown in Fig.~\ref{fig:OBC}. The analysis indicates that both models are consistent with current observational data. At the present level of measurement precision, the differences between the two cases appear to be quite small.

\section{Simulation with Monte Carlo Markov Chain} \label{sec:MCMC}
Following the preparation of the theoretical model and observational dataset, we use a MCMC approach to constrain the Lorentz-violating parameter $\ell$. Our analysis utilizes the Python package emcee \cite{Foreman-Mackey:2012any}, which is well-suited for exploring the correlation of the high-dimensional parameter space. By combining all three types of S2 data mentioned in section  \ref{sec: observable dataset} into a single likelihood, we can use MCMC to find a constraint for $\ell$ based on the current observational data.

\subsection{Analysis of MCMC}
Our MCMC simulation samples a 14-dimensional parameter space,
\begin{equation}
\label{eq:parameter}
\Big\{ M,R_{0},a,e,\iota,\omega,\Omega,t_{\text{apo}},\mathsf{x_0},\mathsf{y_{0}},\mathsf{v_{x0}},\mathsf{v_{y0}},\mathsf{v_{z0}}, \ell \Big\}.
\end{equation}
These parameters are categorized into four groups, including the  the central mass and its distance from Earth $\{M, R_0\}$, and the six Keplerian orbital elements $\{a, e, \iota, \omega, \Omega, t_{\text{apo}}\}$ that define the geometry of the S2's trajectory. We also incorporate five reference-frame terms $\{\mathsf{x_0}, \mathsf{y_{0}}, \mathsf{v_{x0}}, \mathsf{v_{y0}}, \mathsf{v_{z0}}\}$ to correct for coordinate offsets and linear drifts relative to the center black hole Sgr~A*. The final component of the parameter space is the Lorentz-violating term $\{\ell\}$, which is the central focus of our analysis.

Therefore, the total log-likelihood $\log \mathcal{L}$ is the sum of three independent components,
\begin{equation}
\log \mathcal{L} =\log \mathcal{L}_{AP}+\log \mathcal{L}_{V_R}+\log \mathcal{L}_{\Delta\phi}.
\end{equation}
The term $\log \mathcal{L}_{AP}$ accounts for the 145 astrometric positions of S2,
\begin{equation} \label{eq:AP}
\log \mathcal{L}_{AP} = -\frac{1}{2}\sum_{i} \frac{(X_{\text{obs}}^i-X_{\text{the}}^i)^2}{(\sigma^i_{X,\text{obs}})^2}-\frac{1}{2}\sum_{i} \frac{(Y_{\text{obs}}^i-Y_{\text{the}}^i)^2}{(\sigma^i_{Y,\text{obs}})^2} .
\end{equation}
The term $\log \mathcal{L}_{V_R}$ evaluates the 44 radial velocity points:
\begin{equation} \label{eq:VR}
\log \mathcal{L}_{V_R} = -\frac{1}{2}\sum_{i} \frac{(V_{R,\text{obs}}^i-V^i_{R,\text{the}})^2}{(\sigma^i_{V_{R, \text{obs}}})^2}.
\end{equation}
The final term $\log \mathcal{L}_{\Delta\phi}$ constrains the orbital precession using the measured factor
\begin{equation} \label{eq:lP}
\log \mathcal{L}_{\Delta\phi} = -\frac{1}{2}\frac{(f_{\rm sp,\text{obs}}-f_{\rm sp,\text{the}})^2}{\sigma^2_{f_{\rm sp,\text{obs}}}} .
\end{equation}
The variables in the likelihood functions \eqref{eq:AP}--\eqref{eq:lP} are constructed by combining the observational datasets with our gravitational model. Specifically, $X^i_{\text{obs}}$, $Y^i_{\text{obs}}$, and $V^i_{R,\text{obs}}$ represent the measured sky-plane coordinates and radial velocities of the S2 star at each each observational epoch $i$. Their theoretical counterparts, $X_{\text{the}}^i$, $Y_{\text{the}}^i$, and $V^i_{R, \text{the}}$, are obtained by numerically integrating the geodesic equations \eqref{eq:second_order} based on the Schwarzschild-like metric \eqref{eq:kr_metric_sec2}. For the additional constraint from the orbital precession, we compare the observationally determined factor $f_{\text{sp,obs}}$ from Eq.~\eqref{eq:deltaPHI} with its theoretical prediction $f_{\text{sp,the}}$. Here, the theoretical factor is defined as the ratio $f_{sp, \text{the}} \equiv \Delta\phi/\Delta\phi_{GR}$ where the anomalous precession $\Delta\phi$ represents the total precession per orbit as evaluated in Eq.~\eqref{eq:kr_orbit_period}. In all likelihood expressions, $\sigma$ represents the statistical uncertainty for each data point.

\subsection{Results}
We perform MCMC simulation over a 14-dimensional parameter space using two types of different prior distributions, {i.e., the uniform prior and Gaussian prior, as addressed in \cite{DellaMonica:2021xcf}. On the one hand, comparing against GR, the gravitational theory is modified by the Kalb--Ramond field, so in principle there is no prior knowledge for this theory and all orbital parameters may deviate from their GR predictions. Thus, we firstly adopt uniform prior, which are listed in the second and third columns of Table~\ref{tab:table01}. On the other hand, in order to check if the constraints significantly depend on the prior, we also adopt the Gaussian prior which are listed in the last two columns of Table~\ref{tab:table01}. Noted that the choice of Gaussian prior are centered on the best-fit values of parameters provided by the Gravity Collaboration \cite{GRAVITY:2020gka}, due to the fact that no strong Lorentz violation is expected.}  For the Lorentz-violating parameter $\ell$, we apply a uniform prior within $[-0.9, 0.9]$ for both cases. This setup allows us to test the sensitivity of our results to different prior assumptions.

\begin{table}[ht]
\centering
\setlength{\tabcolsep}{10pt}
\renewcommand{\arraystretch}{1.2}
\begin{tabular}{|l|c|c|c|c|}
\hline
\multirow{2}{*}{Parameter} & \multicolumn{2}{c|}{Uniform prior} & \multicolumn{2}{c|}{Gaussian prior} \\
\cline{2-5}
 & Start & End & $\mu$ & $\sigma$ \\
\hline
$a$ (mas)        & 123       & 133    & 125.058   & 0.041     \\
$e$              & 0.5       & 0.9    & 0.884649  & 0.000066  \\
$M$ ($10^6 M_{\bigodot}$) & 2.0 & 6.5   & 4.261    & 0.0012   \\
$t_{\text{apo}}$ (yr)       & 1994.0    & 1994.8   & 1994.2913 & 0.0016   \\
$\mathsf{x_0}$ (mas)      & -1        & 1         & -0.9     & 0.14     \\
$\mathsf{y_0}$ (mas)      & -1        & 1         & 0.07     & 0.12     \\
$\iota$ ($^\circ$)   & 132.0     & 136.0    & 134.567  & 0.033    \\
$\omega$ ($^\circ$)  & 65.0      & 68.7     & 66.263   & 0.031    \\
$\Omega$ ($^\circ$)  & 217.0     & 230.5    & 228.171  & 0.031    \\
$R_0$ (kpc)          & 4         & 12       & 8.2467   & 0.0093   \\
$\mathsf{v_{x0}}$ (mas/yr)   & -1.0      & 1.0      & 0.080    & 0.010    \\
$\mathsf{v_{y0}}$ (mas/yr)   & -1.0      & 1.0      & 0.0341   & 0.00096  \\
$\mathsf{v_{z0}}$ (km/s)     & -50       & 50       & -1.6     & 1.4      \\
\hline
$\ell$              & \multicolumn{4}{c|}{Uniform prior within $[-0.9,0.9]$} \\
\hline
\end{tabular}
\caption{Two different sets of prior used for our MCMC analysis. The Gaussian prior data are taken from \cite{GRAVITY:2020gka}. In this table, $kpc$ is the kiloparsec, $mas$ is the milliarcsecond, $\circ$ is the degree, and $yr$ is the year.}
\label{tab:table01}
\end{table}

\begin{figure} [ht]
{\centering
\includegraphics[width=6.5in]{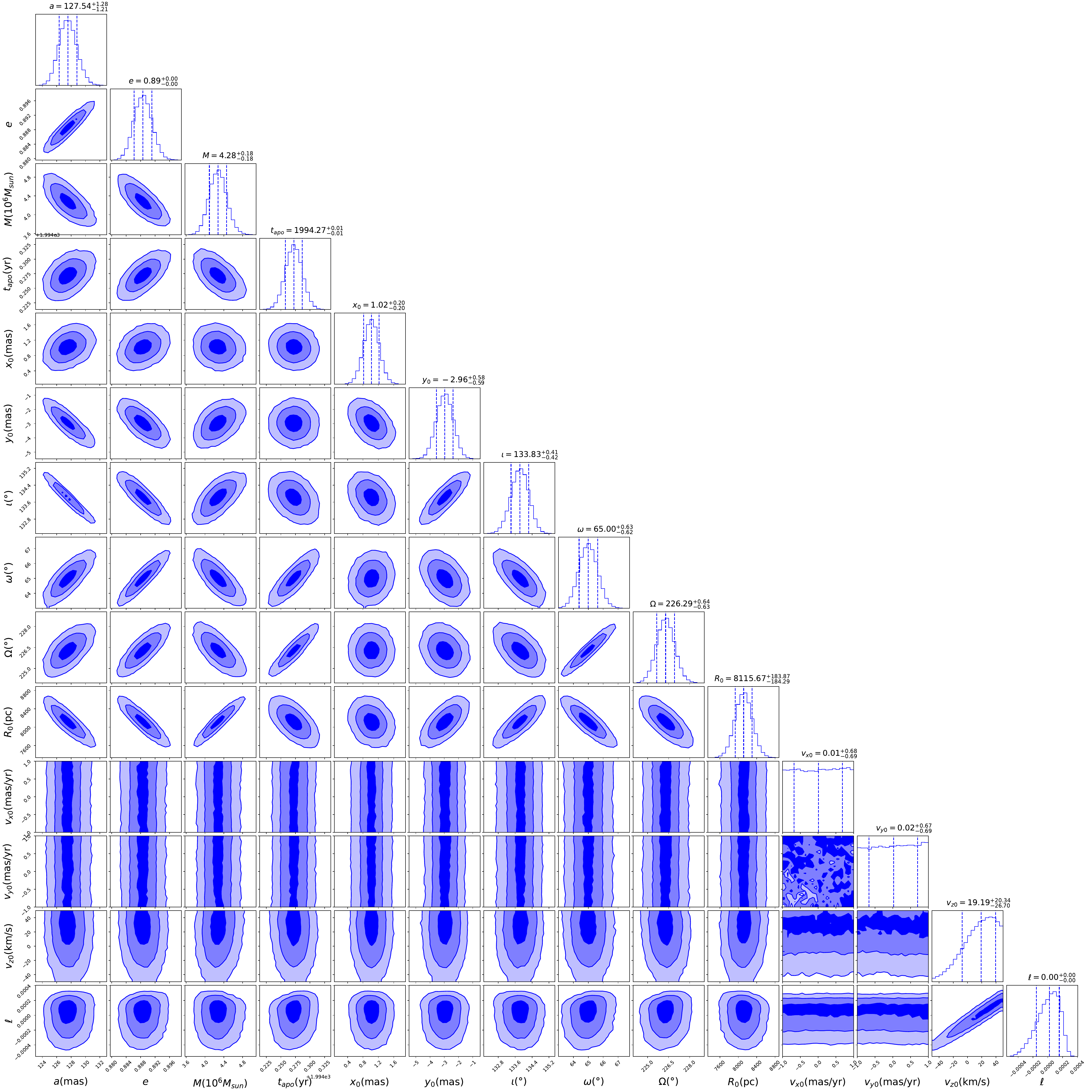} \hspace{0.5cm}
   \caption{ The figure shows the posterior distributions of the orbital parameters of  S2 and the LSB parameter of the Kalb--Ramond gravity gravity theory with uniform priors for the orbital parameters. The dashed lines marks values of the best fit and $1\sigma$ confidence levels.}   \label{fig:kr_uni}}
\end{figure}

\begin{figure} [ht]
{\centering
\includegraphics[width=6.5in]{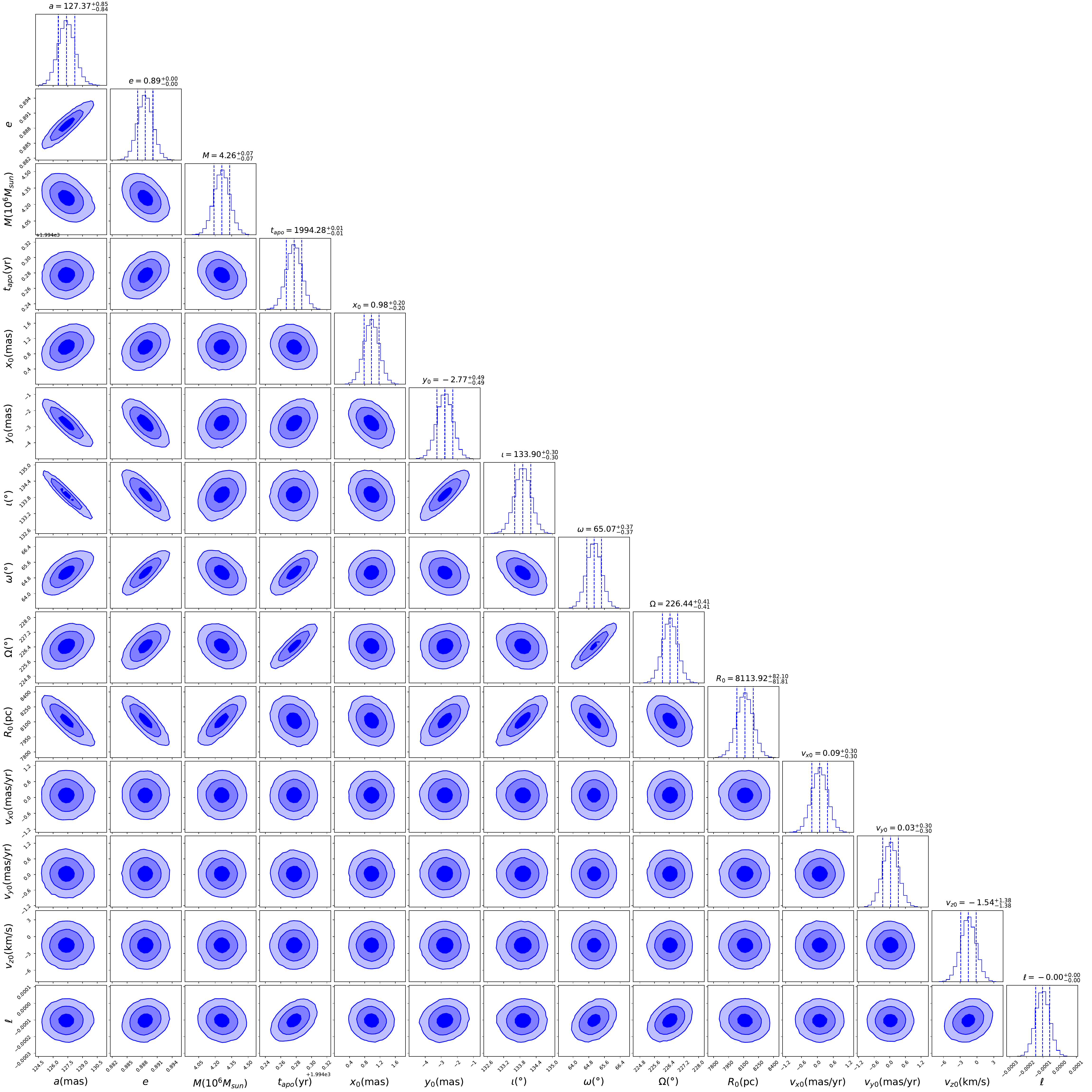} \hspace{0.5cm}
   \caption{ The figure shows the posterior distributions of the orbital parameters of  S2 and the LSB parameter of the Kalb--Ramond gravity theory with Gaussian priors for the orbital parameters. The dashed lines marks values of the best fit and $1\sigma$ confidence levels.}   \label{fig:kr_Gau}}
\end{figure}

The posterior distributions of the 14-parameter space are shown in Fig.~\ref{fig:kr_uni} and Fig.~\ref{fig:kr_Gau}, which are the results of the uniform distribution and the Gaussian distribution, respectively.
In these corner plots, the diagonal panels display the one-dimensional marginal probability density functions, while the color intensity of the off diagonal contours from the inside to the outside indicates the correlation of parameter pairs at the $1\sigma, 2\sigma,$ and $3\sigma$ confidence levels. The corresponding best-fit values and their $1\sigma$ uncertainties are summarized in Table~\ref{tab:table02}. {Our results show that the orbital parameters are well-constrained for both prior choices. By comparing the results in detail, we see that the best-fit values of  the central mass and its distance from Earth $\{M, R_0\}$, and the six Keplerian orbital elements $\{a, e, \iota, \omega, \Omega, t_{\text{apo}}\}$ are  almost consistent for the two sets of priors, but the Gaussian prior shifts the best-fit values of the reference-frame terms $\{\mathsf{x_0}, \mathsf{y_{0}}, \mathsf{v_{x0}}, \mathsf{v_{y0}}, \mathsf{v_{z0}}\}$ from those for the uniform prior, especially the velocity drifts of the reference frame.
Regarding the Lorentz-violating parameter $\ell$, the constraints remain significantly consistent regardless of the choice of prior.} Under the uniform prior, we obtain
\begin{equation}\label{eq:bound1}
\ell = {1.60 \times 10^{-5}}^{+1.38 \times 10^{-4}}_{-1.76 \times 10^{-4}} ,
\end{equation}
while under the Gaussian prior yields
\begin{equation}\label{eq:bound2}
\ell = {-1.02 \times 10^{-5}}^{+1.26 \times 10^{-4}}_{-1.19 \times 10^{-4}} .
\end{equation}
To verify the stability of these results, we also test narrower prior ranges, such as $\ell \in [-10^{-2}, 10^{-2}]$ and $\ell \in [-10^{-3}, 10^{-3}]$, and found the resulting posterior distributions consistent.

\begin{table}[ht]
\centering
\setlength{\tabcolsep}{10pt}
\renewcommand{\arraystretch}{1.2}
\begin{tabular}{|l|c|c|}
\hline
Parameter & best-fit value (Uniform Prior) & best-fit value (Gaussian Prior) \\
\hline
$a$ (mas)           & $127.54^{+1.28}_{-1.21}$   & $127.37^{+0.85}_{-0.84}$ \\
$e$                 & $0.89 \pm 0$   &  $0.89 \pm 0$  \\
$M$ ($10^6 M_{\bigodot}$)  &  $4.28 \pm 0.18$          & $4.26 \pm 0.07$  \\
$t_{\text{apo}}$ (yr)       & $1994.27 \pm 0.01$   & $1994.28 \pm 0.01$  \\
$\mathsf{x_0}$ (mas)       &  $1.02 \pm 0.20$   & $0.98 \pm 0.20$ \\
$\mathsf{y_0}$ (mas)     &   $-2.96 ^{+0.58}_{-0.59}$   & $-2.77 \pm 0.49$ \\
$\iota$ ($^\circ$)     & $133.83^{+0.41}_{-0.42}$     & $133.90\pm 0.30$  \\
$\omega$ ($^\circ$)    & $65.00^{+0.63}_{-0.62}$  & $65.07 \pm 0.37$ \\
$\Omega$ ($^\circ$)    & $226.29^{+0.64}_{-0.63}$ & $226.44 \pm 0.41$  \\
$R_0$ (pc)     & $8115.67^{+183.87}_{-184.29}$ &$8113.92^{+82.10}_{-81.81}$ \\
$\mathsf{v_{x0}}$ (mas/yr)      & $0.01 ^{+0.68}_{-0.69}$  & $0.09 \pm 0.30 $   \\
$\mathsf{v_{y0}}$ (mas/yr)      & $0.02^{+0.67}_{-0.69}$  & $0.03 \pm 0.30 $ \\
$\mathsf{v_{z0}}$ (km/s)        & $19.19 \pm ^{+20.34}_{-26.70}$  & $-1.54 \pm 1.38$  \\
$\ell$          & ${1.60 \times 10^{-5}}^{+1.38 \times 10^{-4}}_{-1.76 \times 10^{-4}}$   & ${-1.02 \times 10^{-5}}^{+1.26 \times 10^{-4}}_{-1.19 \times 10^{-4}}$   \\
\hline
\end{tabular}
\caption{The best-fit values of the parameters of the orbital
model of S2 with two different priors. The uncertainties are given at the $1\sigma$ level.
\label{tab:table02} }
\end{table}

It is noted that recent studies of this metric suggest that EHT data constrain this parameter to approximately
$-0.185022 \leq \ell \leq 0.060938$ \cite{Junior:2024vdk,Junior:2024ety}. Our current study shows that the orbital dynamics of the S2 star provide a much more sensitive probe. The bounds we obtained in Eq.~\eqref{eq:bound1} and \eqref{eq:bound2} are three orders of magnitude tighter than those from current black hole imaging. This demonstrates that we can get much more precise limits on $\ell$ by tracking S2's orbit than by using current black hole imaging from the EHT.
Furthermore, a recent analysis on EMRI waveform with a one-year observation period indicates that the the  effect of LSB in the current model becomes detectable by LISA and its detection error from Fisher information matrix approach is  $|\Delta \ell| \sim 10^{-5}$ \cite{Xia:2025yzg}, which is better than our $\sim 10^{-4} $.
Additionally, while the constraints on $\ell$ in Eq.~\eqref{eq:bound1} and Eq.~\eqref{eq:bound2} are not as tight as those obtained from Solar System planetary data \cite{Yang:2023wtu,Junior:2024ety,AraujoFilho:2025jcu}, such a difference is anticipated. The precision of S2 star observations is limited by current instrument resolution and the complex dynamical environment of the Galactic Center. But the significance of the S2 star is that it provides a new way to verify the theory of gravity and its extensions. Unlike the weak gravitational environment of the Solar System, the Galactic Center allows us to test LSB in an extremly strong gravitational region near the supermassive black hole. Our results demonstrate that even though the current limits of observational precision, stellar dynamics is sensitive enough to give meaningful constraints on the LSB in the related GR modifications.

\section{Conclusion and discussion} \label{sec:conclusion}

In this study, we investigated the feature of the Kalb-Ramond field by analyzing the orbital motion of the S2 star near the Galactic Center. By combining astrometric and spectroscopic data with the theoretical orbit derived from the geodesic equations, we performed a 14-dimensional MCMC simulation to constrain the Lorentz-violating parameter $\ell$. Using two different sets of priors, we obtained the following constraints at the $1\sigma$ confidence level,
\begin{align}
\text{uniform prior}: \ell = {1.60 \times 10^{-5}}^{+1.38 \times 10^{-4}}_{-1.76 \times 10^{-4}},\
~~\text{Gaussian prior}: \ell = {-1.02 \times 10^{-5}}^{+1.26 \times 10^{-4}}_{-1.19 \times 10^{-4}},
\end{align}
These results indicate that at $1\sigma$ confidence level, no significant deviation from GR within current observational uncertainties.

An important finding of this work is that the constraints derived from S2 star are much more stringent than those obtained from black hole shadow imaging. While recent studies using EHT data for Kalb--Ramond gravity show a constraint of roughly $\ell < \mathcal{O}(10^{-2})$ \cite{Junior:2024ety}, our results improve this limit by nearly three orders of magnitude. Similarly,  considering the shadow size of Sgr~A* along with stellar dynamics priors, Ref. \cite{Lobos:2026ysx} placed a conservative bound of $\ell \lesssim 0.19$. Whereas the S2 star's orbital dynamics in the current study  constrain $\ell$ at the $\mathcal{O}(10^{-5})$ level. This result suggests that studying the long-term dynamics of the S2 star provides a preciser way to detect the LSB induced by Kalb--Ramond field than current observations of the shadow of Sgr~A*.
 On the other hand, although Solar System tests (e.g., Shapiro's time-delay) provide the most stringent limits on Lorentz violation in the Kalb--Ramond gravity \cite{Yang:2023wtu, AraujoFilho:2025jcu}, the present work probes a completely different physical regime characterized by strong gravitational fields near the SMBH. This highlights the importance of Galactic Center stellar dynamics as an independent and complementary probe of fundamental physics.

\begin{figure} [ht]
{\centering
\includegraphics[width=2.25in]{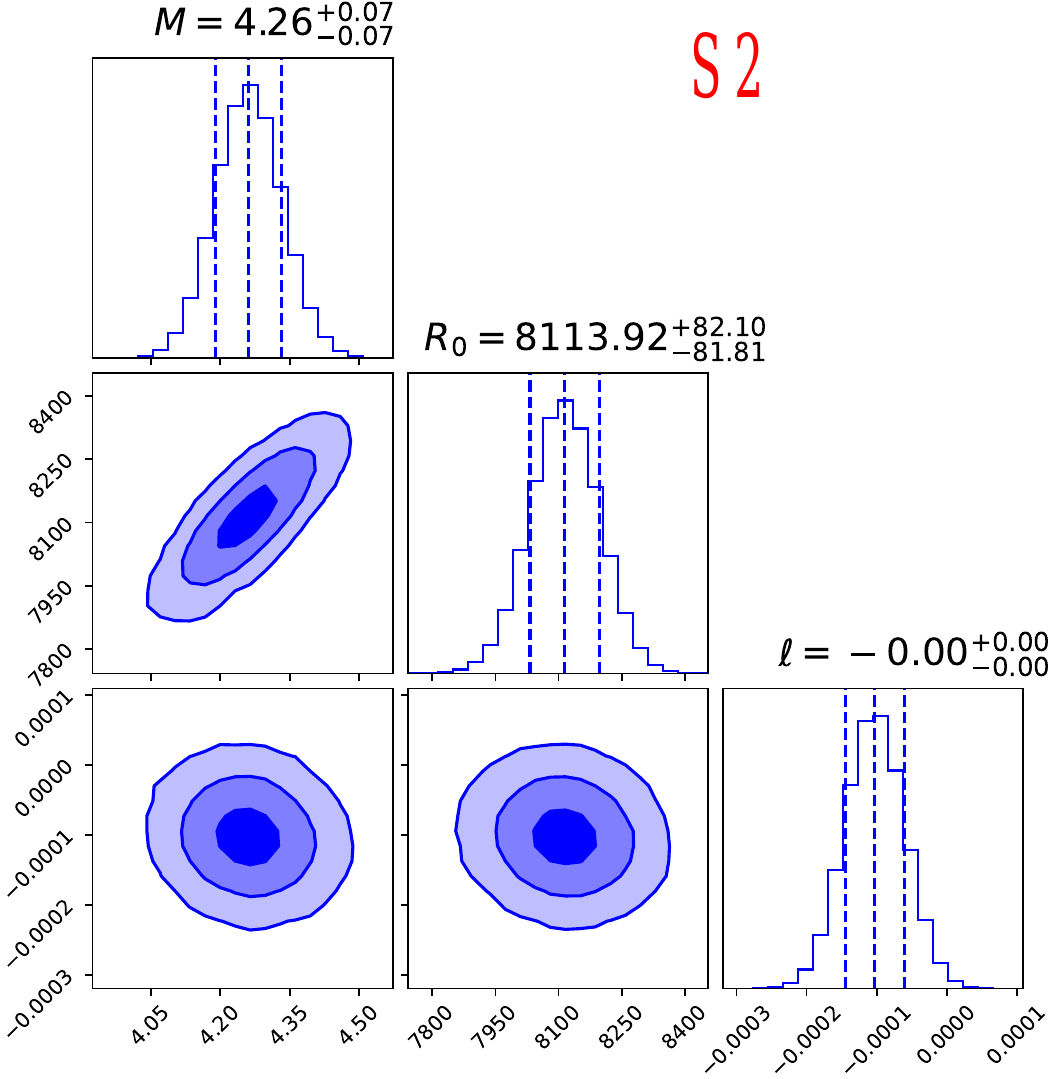}\hspace{0.1cm}
\includegraphics[width=2.25in]{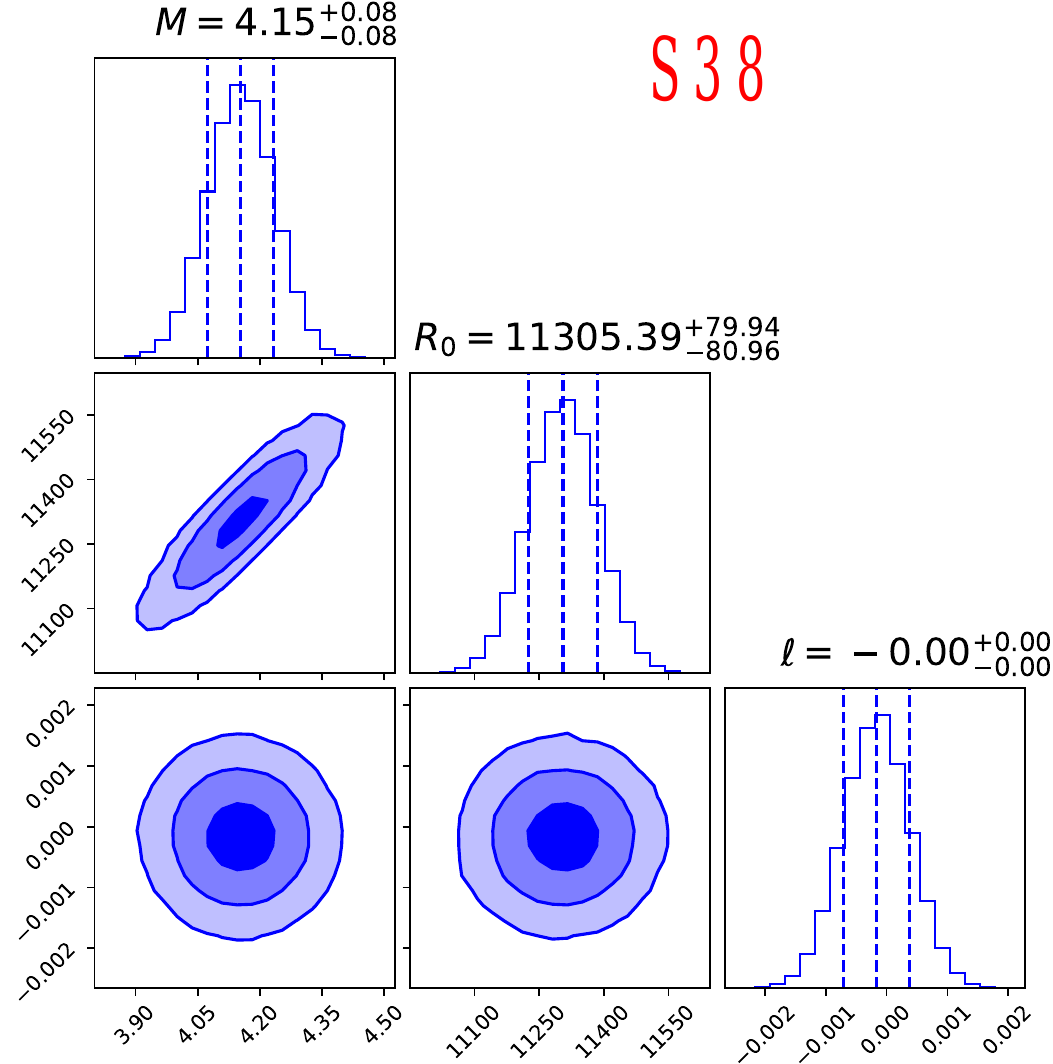}\hspace{0.1cm}
\includegraphics[width=2.25in]{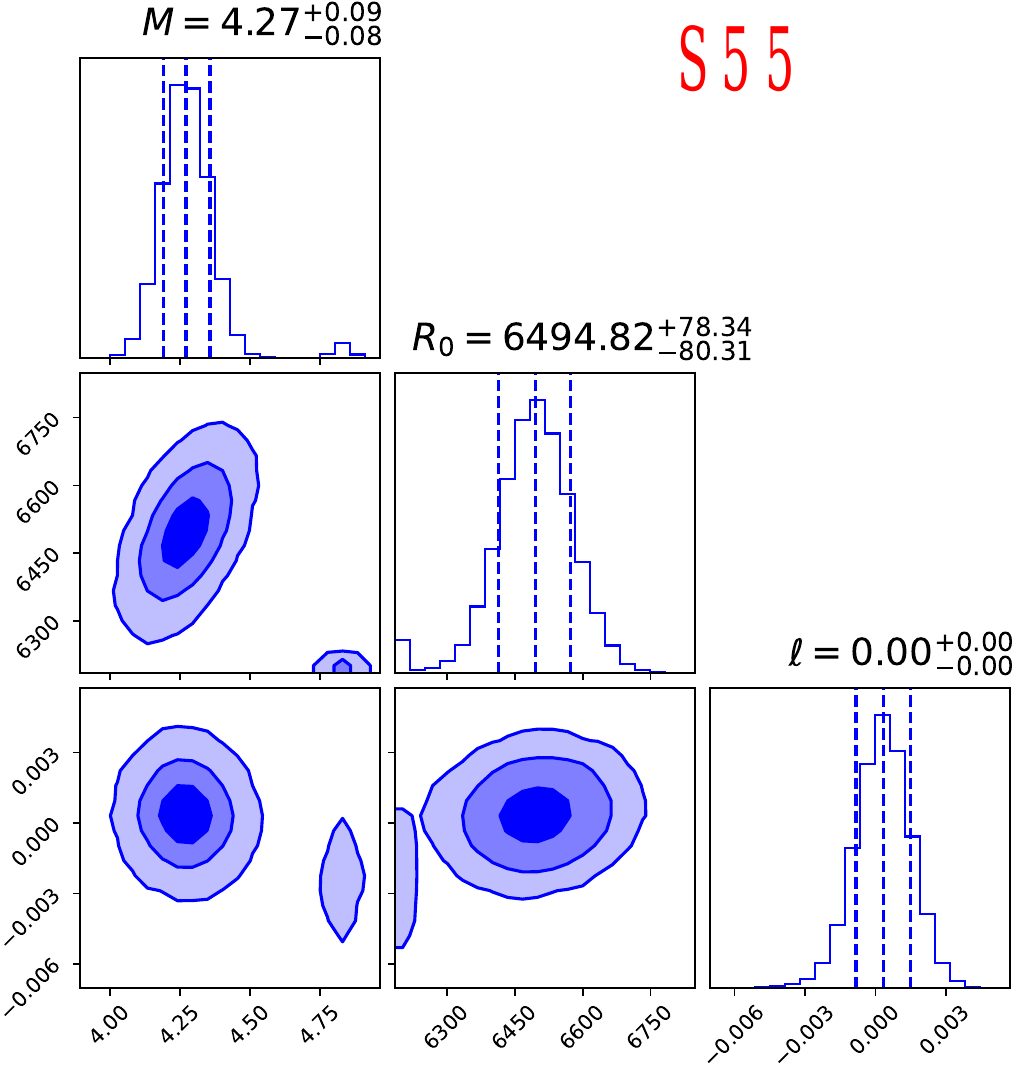}
\caption{ The posterior distributions of the central black hole parameters $\{M, R_0\}$ and the LSB parameter $\ell$, obtained from the MCMC fitting of a single-star. From left to right, the plots illustrate the results of the S2 star, the S38 star, and the S55 star, respectively. The contour lines represent the 1$\sigma$ confidence levels and the best-fit values.}
\label{fig:single}}
\end{figure}

\begin{figure} [ht]
{\centering
\includegraphics[width=2.25in]{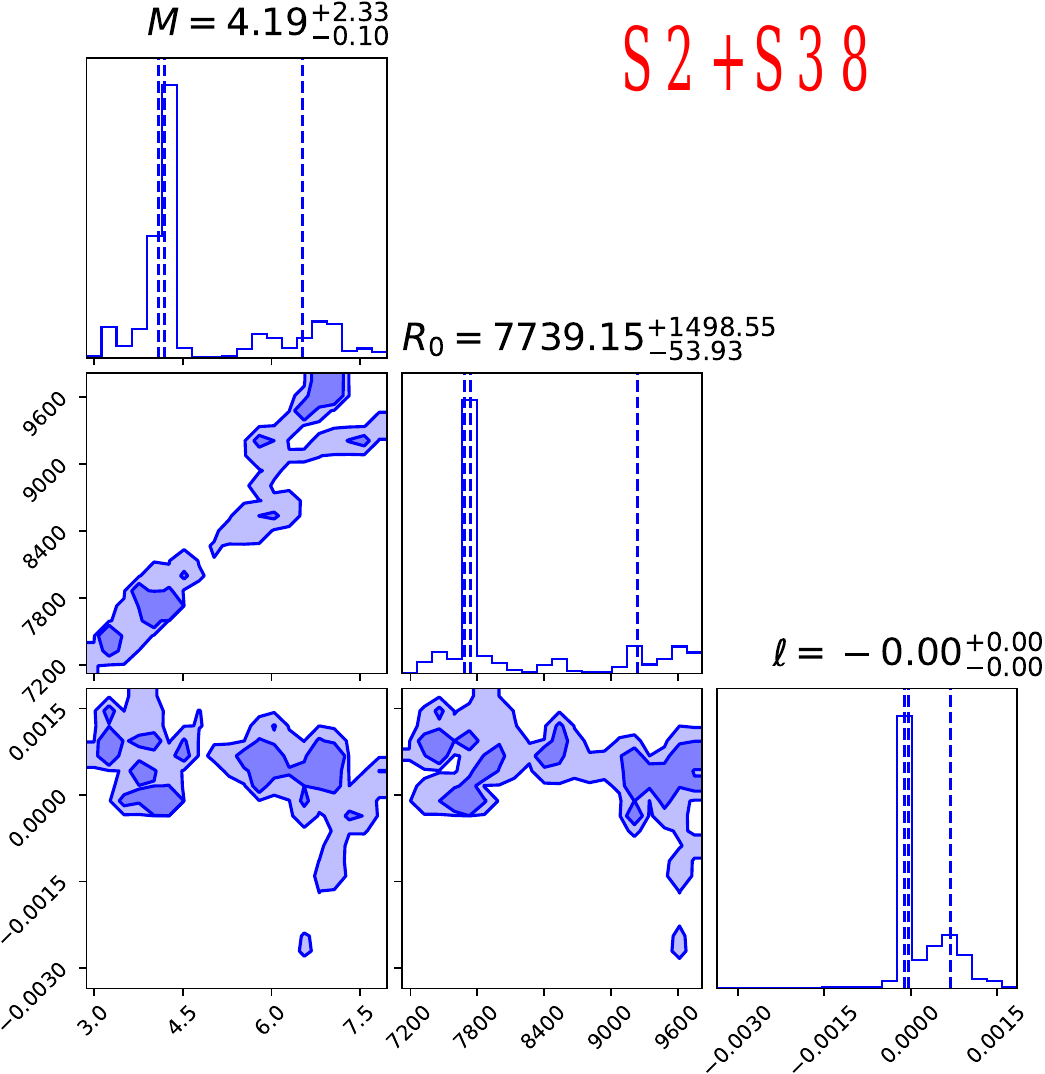}\hspace{0.1cm}
\includegraphics[width=2.25in]{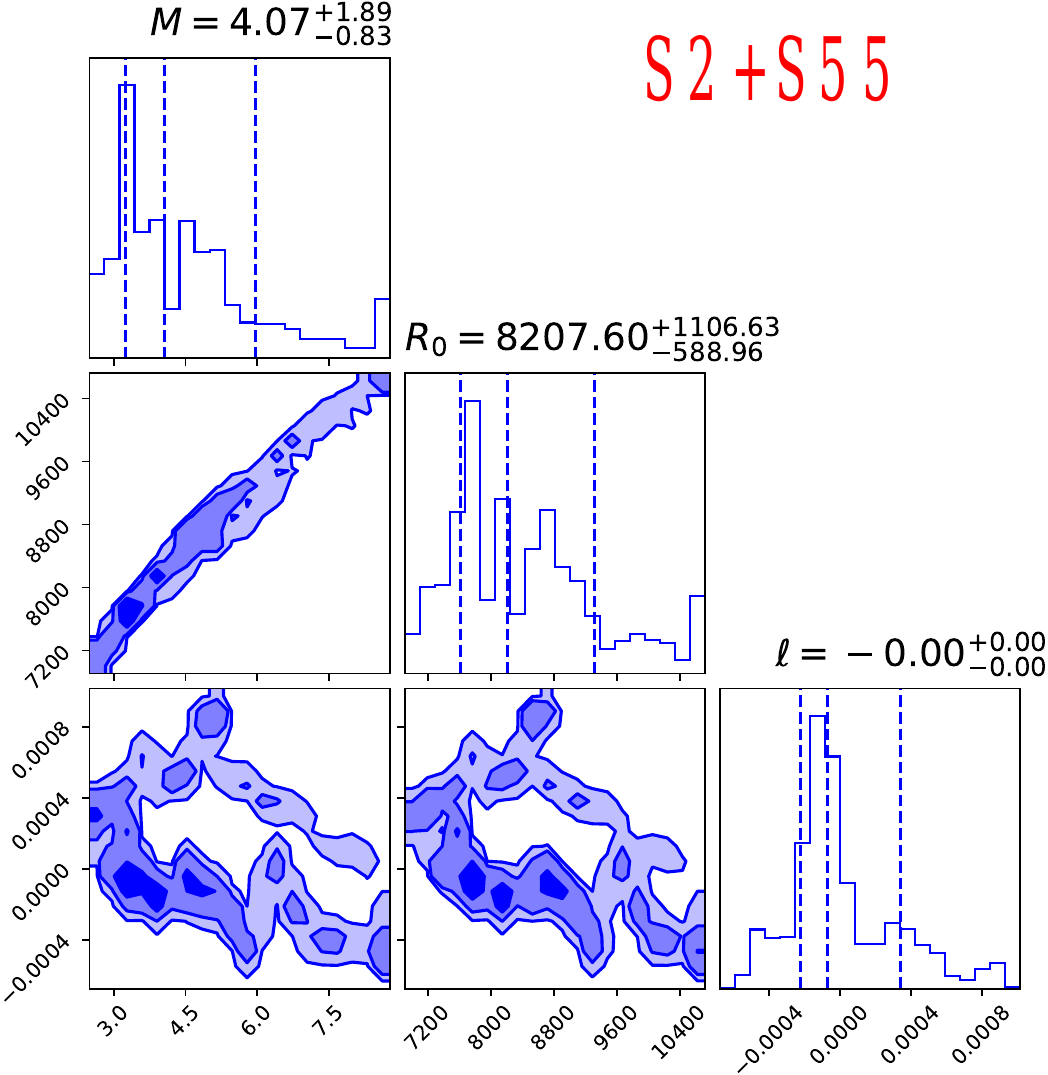}\hspace{0.1cm}
\includegraphics[width=2.25in]{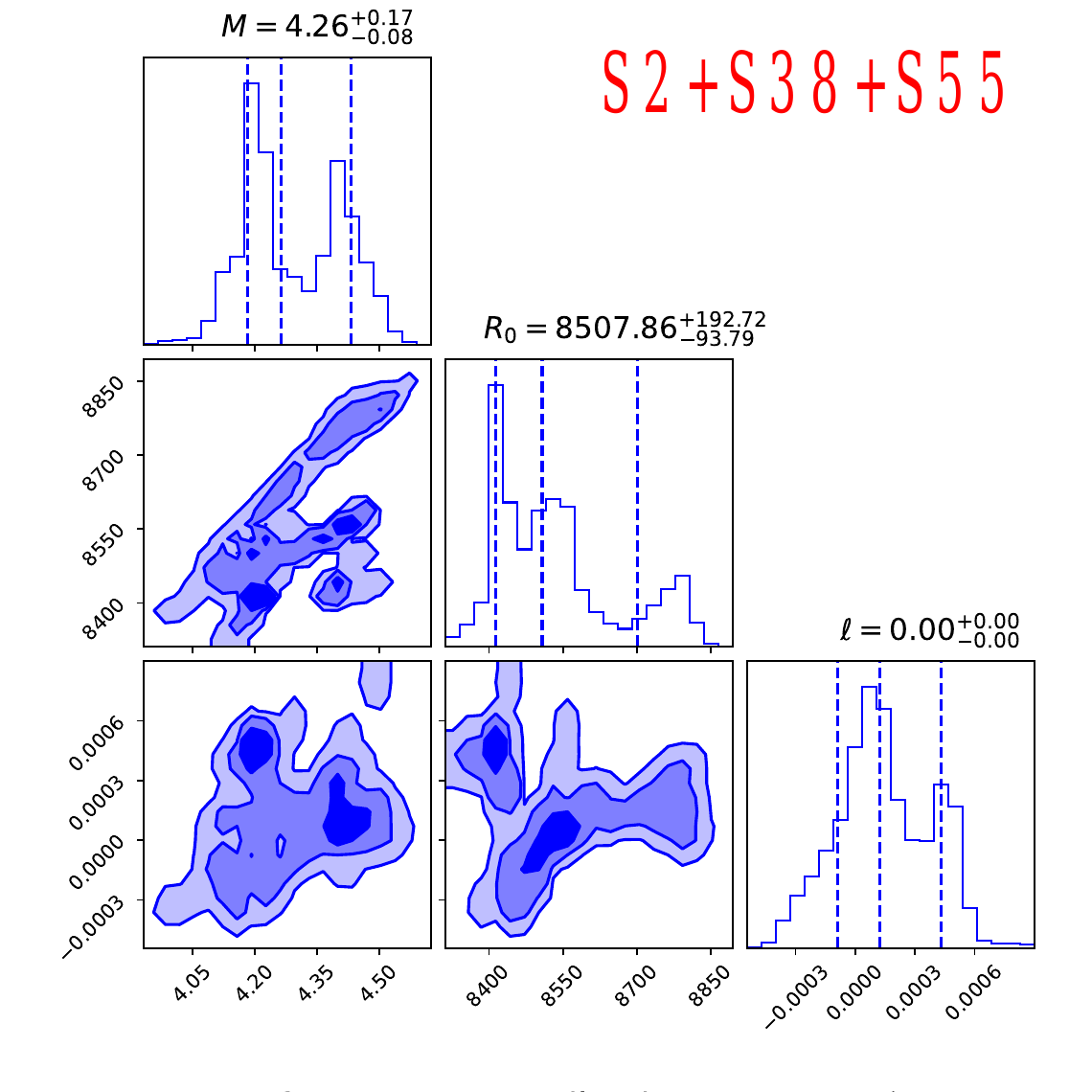}
\caption{ The posterior distributions of the central black hole parameters $\{M, R_0\}$ and the LSB parameter $\ell$, obtained from a multi-stars MCMC simulations.  The plots illustrate the results of different joint fits, corresponding to the S2$+$S38 stars (left), the S2$+$S55 stars (Middle), and the S2$+$S38$+$S55 stars (Right), respectively. The contour lines represent the 1$\sigma$ confidence levels and the best-fit values.}
\label{fig:joint}}
\end{figure}

In the main text, we only focused on the S2 star because it provides the most comprehensive and high-precision available dataset. There could be several ways we can implement to improve these constraints. One primary improvement would be to extend from a single-star analysis to a joint dynamical model that includes multiple S-stars, such as S55 and S38. While integrating stars such as S38 and S55 is theoretically promising for reducing systematic errors and improving prediction accuracy of gravity tests \cite{Shen:2023kkm, jovanovic2023constraints, Pietroni:2022cur, Borka:2022sot,Benisty:2021cmq}, but their currently public orbital data are much less than those of S2 star. Even though, with the same method as described in the simulation with data of S2 star, we also perform the MCMC simulation with single star or joint stars.  It is noted that for a single star, our MCMC simulation is a 14-dimensional parameter space, same as \eqref{eq:parameter}, while for joint stars, we shall stack up all their Keplerian orbital elements. To explicitly exhibit the constraints from the stars, we only extract the posterior distributions of  the central mass and its distance from Earth $\{M, R_0\}$, and the LSB parameter ${\ell}$ in Figs.~\ref{fig:single} and ~\ref{fig:joint}, and their best-fit values are listed in Table~\ref{table03}.  Obviously, S2 star provides the most stringent constraint on  $M$,  $R_0$ and $\ell$ compared against other single-star simulations. This is reasonable because among them, S2 star has shortest orbital period and highest precision observational data  while  S38 and S55 stars both exhibit larger uncertainties, due to  the fewer public data points available for them. In particular, for the LSB parameter $\ell$, all single-star  and multi-stars joint fits constrain its values within the order of $10^{-5}$ as shown in Table~\ref{table03}. It means that although adding S38 and S55 introduces additional orbital parameters and data, the constraint on $\ell$ remains almost stable. This consistency demonstrates that our current constraint on the LSB parameter $\ell$  is robust against the choice of specific S-stars, and the joint fits do not significantly improve over S2 star  because of the current limitations in the public data quality of S38 and S55.
Therefore, with next-generation facilities like GRAVITY+ \cite{abuter2022first} and the ELT \cite{Padovani:2023dxc} coming, the precision of astrometric and spectroscopic data will continue to improve and  more data become available, we expect this advancements in the near future will enable us to test LSB in strong-gravity regions in a preciser way.

\begin{table}[ht]
\centering
\renewcommand{\arraystretch}{1.2}
\begin{tabular}{|c|c|c|c|}
\hline
& $S2$ & $S38$  & $S55$    \\ \hline
 $M$ ($10^6 M_{\bigodot}$)  & $4.26 \pm 0.07$ & $4.15 \pm 0.08$  &  $4.27^{+0.09}_{-0.08}$   \\ \hline
$R_0$ (pc) & $8113.92^{+82.10}_{-81.81}$ & $11305.39^{+79.97}_{-80.96}$  &  $6494.82^{+78.34}_{-80.31}$   \\ \hline
 $\ell$ & ${-1.02 \times 10^{-5}}^{+1.26 \times 10^{-4}}_{-1.19 \times 10^{-4}}$  & ${-1.61 \times 10^{-5}}^{+5.45 \times 10^{-4}}_{-5.34 \times 10^{-4}}$   &  ${3.33 \times 10^{-5}}^{+1.15 \times 10^{-4}}_{-1.16 \times 10^{-4}}$  \\ \hline
 \hline
 & $S2 + S38$  & $S2 + S55$  & $S2 + S38 + S55$ \\ \hline
$M$ ($10^6 M_{\bigodot}$)  & $4.19^{+2.33}_{-0.10}$ & $4.07^{+1.89}_{-0.83}$ & $4.26^{+0.17}_{-0.08}$ \\ \hline
$R_0$ (pc) & $7739.15^{+1498.55}_{-53.93}$ & $8734.80^{+1106.63}_{-715.29}$ & $8507.86^{+192.72}_{-93.79}$ \\ \hline
$\ell$ & ${-3.52 \times 10^{-5}}^{+7.25 \times 10^{-4}}_{-7.19 \times 10^{-4}}$ & ${-6.88 \times 10^{-5}}^{+4.09 \times 10^{-4}}_{-1.56 \times 10^{-4}}$ & ${1.22 \times 10^{-5}}^{+3.10 \times 10^{-4}}_{-2.10 \times 10^{-4}}$  \\ \hline
\end{tabular}
\caption{The best-fit values and 1$\sigma$ uncertainties of the black hole parameters $(M,R_0)$ and the LSB parameter $\ell$, obtained from the single-star and joint multi-stars fits.
\label{table03} }
\end{table}

Another improvement is more than the Schwarzschild-like assumption. Since Sgr A* is rotating, accounting for its spin is necessary for a more realistic description. While a full Kerr-like solution in Kalb--Ramond gravity is still under development, recent progress on slowly rotating Kalb--Ramond black holes \cite{Deng:2025atg} and their tidal dynamics \cite{Junior:2026upy} provides a much better starting point. Incorporating these solutions into stellar orbit modeling will be a promising work for our future research.

\begin{acknowledgments}
We appreciate Dr. Qiang Wu for helpful corresponding. This work is partly supported by Natural Science Foundation of China under Grants No.12375054 and 12175192, the Postgraduate Research $\&$ Practice Innovation Program of
Jiangsu Province under Grant No. $\text{KYCX25\_3924}$, and Yangzhou Science and Technology Planning Project in Jiangsu Province of China (YZ2025233).
\end{acknowledgments}

\providecommand{\href}[2]{#2}\begingroup\raggedright\endgroup

\end{document}